\documentclass[prb,aps,tightenlines,twocolumn,
floatfix]{revtex4}
\usepackage{graphicx}
\usepackage{amssymb}
\usepackage{amsmath}
\usepackage{bm}
\usepackage{colordvi}
\usepackage{mathrsfs}

\begin{document}
\title{Comment on ``Floquet spin states in graphene under ac-driven spin-orbit
  interaction''} 
\author{Y. Zhou}
\author{M. W. Wu}
\thanks{Author to whom correspondence should be addressed}
\email{mwwu@ustc.edu.cn.}
\affiliation{Hefei National Laboratory for Physical Sciences at
  Microscale and Department of Physics, University of Science and
  Technology of China, Hefei, Anhui, 230026, China}

\date{\today}



\maketitle

Recently, L\'opez {\em et al.}\cite{Schliemann_ac_Rashba} applied the
Magnus-Floquet expansion\cite{Blanes_PR} (MFE) to the quasi-energy spectrum and
the related dynamical features in monolayer graphene under the periodically
Rashba spin-orbit interaction.
They suggested this approach to be an efficient tool to deal with 
time-dependent problems and further claimed that the results from this approach 
converge faster than those from the standard Floquet-Fourier
approach\cite{Shirley_65,Hanggi_rev,Syzranov_08,Oka_cur,Zhou_THz,Calvo_APL} 
for a weak field strength. 
In the following, we will demonstrate that many results in that paper are in
fact beyond the convergence domain of the MFE and hence are incorrect. 

\begin{figure}[tbp]
  \begin{center}
    \includegraphics[width=6cm]{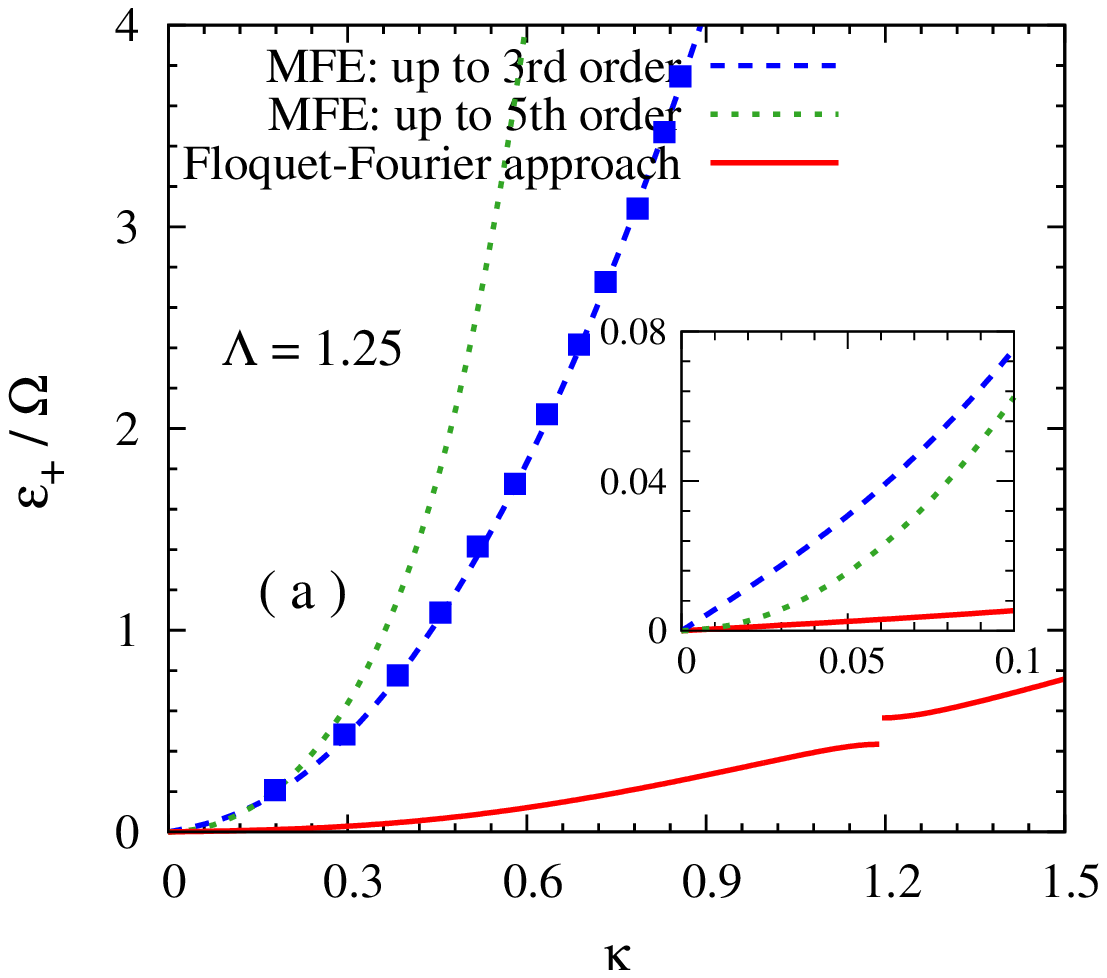}
    \includegraphics[width=6cm]{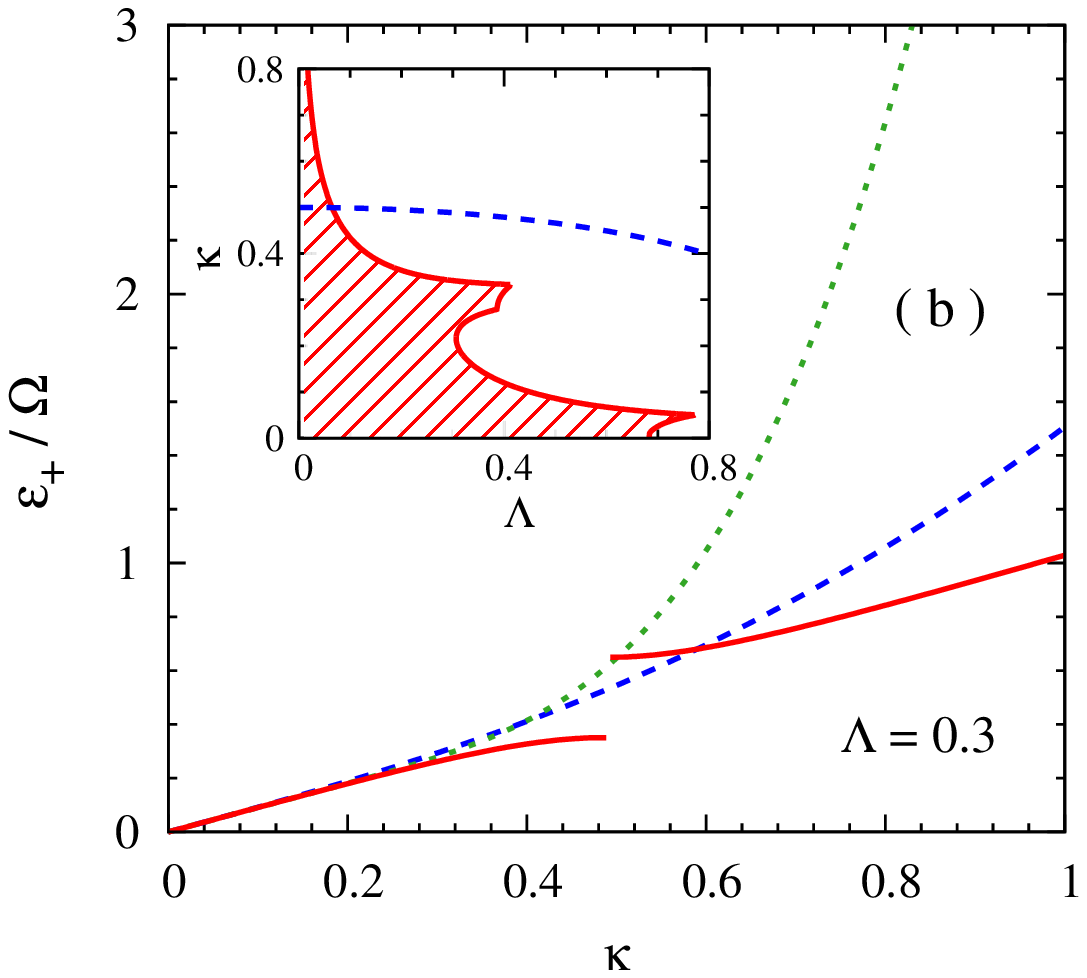}
  \end{center}
  \caption{ (Color online) The quasi-energies $\epsilon_{+}$ from the MFE up to
    third (blue dashed curves) and fifth (green dotted curves) orders 
as well as those from
    the Floquet-Fourier approach (red solid curves) for $\Lambda=1.25$ (a)
 and  $0.3$ (b). The blue squares in (a) are taken from  
 Fig.~1 in Ref.~\onlinecite{Schliemann_ac_Rashba}. 
    The region for $\kappa<0.1$ in (a) is enlarged in its inset. 
    In the inset of (b), the convergence regime of the MFE (the shadow area)
and the momentum corresponding to the first 
    quasi-energy gap as function of $\Lambda$ (blue dashed curve) are 
plotted. 
  }
  \label{fig_com} 
\end{figure}

We first plot the quasi-energies
of the lowest conduction-like Floquet band $ \varepsilon_{+}$, taken from
Fig.~1 in Ref.~\onlinecite{Schliemann_ac_Rashba}, as blue squares in 
Fig.~\ref{fig_com}(a). They are
obtained via the MFE up to the third order:\cite{Schliemann_ac_Rashba,Blanes_PR}
\begin{equation}
  \varepsilon_{+}^{{\rm up\ to}\ (3)}= \Omega \kappa \sqrt{16\kappa^2\Lambda^2
    +(\Lambda^2-1)^2},
  \label{MFE_3}
\end{equation} 
where $\kappa=v_{\rm F} k/\Omega$ and $\Lambda=\lambda_{R}/\Omega$, with
$\lambda_{R}$ and $\Omega$ being the magnitude and frequency of the ac-driven
Rashba spin-orbit coupling.
Since the value of $\Lambda$ was not given in that paper, we 
fit their results with Eq.~(\ref{MFE_3}). 
The best fitting gives $\Lambda=1.25$, with the corresponding results plotted as 
blue dashed curve.
It is seen that no quasi-energy gap opens in the whole momentum regime.
This behavior is very much different from that in graphene irradiated by a laser
field.\cite{Syzranov_08,Oka_cur,Zhou_THz,Calvo_APL}  
However, the Hamiltonian in this system is similar to the latter one. 
This can be seen more clearly by transforming the Hamiltonian
given by Eq.~(11) in Ref.~\onlinecite{Schliemann_ac_Rashba} into
the basis set formed by the eigenvectors of its time-independent term. The
Hamiltonian becomes
\begin{equation}
  \tilde{h}_{-}(k,t)= v_F k {\sigma}_z - \lambda_{R}\cos\Omega t (I+{\sigma}_x),
  \label{h_change}
\end{equation}
with $I$ and ${\bm{\sigma}}$ being the identity and Pauli matrices,
respectively. By comparing this equation with Eqs.~(B3) and (B4) in
our work,\cite{Zhou_THz} one can find $\tilde{h}_{-}(k,t)$ shares the identical
quasi-energy spectrum with that in graphene under a linearly polarized laser
field with the momentum perpendicular to the direction of the laser field.
In the latter case, the previous
investigations\cite{Syzranov_08,Zhou_THz,Calvo_APL} have demonstrated that 
quasi-energy gaps appear at nonzero momentum [a typical case can be seen in Fig.~3(b) in
Ref.~\onlinecite{Zhou_THz}]. 
This indicates that the quasi-energy spectrum reported by L\'opez 
{\em et al.}\cite{Schliemann_ac_Rashba} are questionable.

We further calculate the corresponding quasi-energies of
Eq.~(\ref{h_change}) from the Floquet-Fourier
approach\cite{Shirley_65,Hanggi_rev,Syzranov_08,Oka_cur,Zhou_THz,Calvo_APL} 
[red solid curve in Fig.~\ref{fig_com}(a)]. 
This approach transforms the solving of the time-dependent Schr\"odinger
equation into an eigenvalue problem of the Hamiltonian in the Fourier basis
set. 
By increasing the number of the Fourier mode, one can obtain the quasi-energy
and eigenstates with any degree of accuracy in principle.
Here we calculate the quasi-energies of the system in
Ref.~\onlinecite{Schliemann_ac_Rashba} via this approach using 30 Fourier modes. 
The relative error are verified to be smaller than $10^{-5}$.
From Fig.~\ref{fig_com}(a), one observes that the results from the MFE are
{\em qualitatively} different from the exact results from the Floquet-Fourier approach.
In particular, a gap appears in the quasi-energy spectrum from the
Floquet-Fourier approach, in agreement with the analysis of the Hamiltonian, but
is absent in those from MFE. 
This comparison further confirms that the quasi-energy spectrum in that paper
is incorrect. 
In addition, the results in Figs.~4 and 6 in that paper are based on the same
approach with the same parameter, and hence are questionable as well.

In order to reveal the reason leading to this problem, we plot the
quasi-energies from the MFE up to the fifth order as green
dotted curve. 
The corresponding formula reads\cite{Schliemann_ac_Rashba,Blanes_PR,correct}  
\begin{eqnarray}
  \nonumber
  \varepsilon_{+}^{{\rm up\ to}\ (5)}&=& 
  \frac{\Omega \kappa }{36}[331776\kappa^6\Lambda^2 
  -2304\kappa^4(23\Lambda^4-72\Lambda^2)
  \\ && \mbox{} \hspace{-2.cm}
  +32\kappa^2(1109\Lambda^6-900\Lambda^4-324\Lambda^2)   
  +81(\Lambda^2-2)^4 ]^{\frac{1}{2}}.
  \label{MFE_5}
\end{eqnarray}
One observes that the difference between the results from the MFE up to
the third and fifth orders is considerably grave even at small momentum 
(the region for $\kappa<0.1$ is enlarged in the inset). 
Further calculations show that the convergence criterion, chosen to be 
$|\varepsilon_{+}^{{\rm up\ to}\ (5)}-\varepsilon_{+}^{{\rm up\ to}\ (3)}|
/\varepsilon_{+}^{{\rm up\ to}\ (3)}<10\%$ and
$|\partial_\kappa\varepsilon_{+}^{{\rm up\ to}\ (5)}
-\partial_\kappa\varepsilon_{+}^{{\rm up\ to}\ (3)}|
/\partial_\kappa\varepsilon_{+}^{{\rm up\ to}\ (3)}<10\%$ here,
cannot be satisfied in the whole momentum regime. 
This indicates that the discrepancy between the quasi-energies from the MFE and
the Floquet-Fourier approach is due to the fact that the results 
from the MFE do not converge.

We also plot the quasi-energies $\epsilon_{+}$ from the above approaches for a
weaker field strength $\Lambda=0.3$ in Fig.~\ref{fig_com}(b).
It is seen that $\epsilon_{+}$ from Eq.~(\ref{MFE_3}) agrees well with the exact
one from the Floquet-Fourier approach in the convergence regime of the MFE, but
deviates markedly beyond the convergence regime, e.g., at the
momentum corresponding to the quasi-energy gap. 
This agrees with the discussions presented above.
We further plot the convergence regime of the MFE as well as the momentum
corresponding to the first quasi-energy gap as function of $\Lambda$ in the
inset of Fig.~\ref{fig_com}(b). 
From this inset, one finds that the approach of the MFE only converges for 
small momentum and weak ac field. Moreover, the parameter used in
 Ref.~\onlinecite{Schliemann_ac_Rashba}, i.e.,  $\Lambda=1.25$,
 is well beyond the  convergence regime which is
 in consistence with the results shown in Fig.~\ref{fig_com}(a). 
It is also shown that except for extremely small $\Lambda$, where 
the exact quasi-energy spectrum is very close to the field-free one 
(the relative difference is smaller than 10\%) and 
the quasi-energy gap is still negligible,
the maximum momentum of the convergence regime is always
smaller than the momentum corresponding to the first quasi-energy gap.
This explains why the results from the MFE cannot reproduce
the gap-like behavior.

This work was supported by the National Basic Research Program of China under
Grant No.\ 2012CB922002 and the National Natural Science Foundation of China
under Grant No.\ 10725417.


\begin{thebibliography}{0}
\bibitem{Schliemann_ac_Rashba} A. L\'opez, Z. Z. Sun, and J. Schliemann,
  Phys. Rev. B {\bf 85}, 205428 (2012).
\bibitem{Blanes_PR} S. Blanes, F. Casas, J. A. Oteo, and J. Ros, Phys. Rep. 
  {\bf 470}, 151 (2009).
\bibitem{Shirley_65}  J. H. Shirley, Phys. Rev. {\bf 138}, B979 (1965).
\bibitem{Hanggi_rev} M. Grifoni and P. H\"anggi, Phys. Rep. {\bf 304}, 229
  (1998); S. Kohler, J. Lehmann, and P. H\"anggi, Phys. Rep. {\bf 406}, 379
  (2005). 
\bibitem{Oka_cur} T. Oka and H. Aoki, Phys. Rev. B {\bf 79}, 081406(R) (2009);
  {\it ibid.} {\bf 79}, 169901(E) (2009).
\bibitem{Syzranov_08} S. V. Syzranov, M. V. Fistul, and K. B. Efetov,
  Phys. Rev. B {\bf 78}, 045407 (2008).
\bibitem{Zhou_THz} Y. Zhou and M. W. Wu, Phys. Rev. B {\bf 83}, 245436 (2011). 
\bibitem{Calvo_APL} H. L. Calvo, H. M. Pastawski, S. Roche, and L. E. F. Foa
  Torres, Appl. Phys. Lett. {\bf 98}, 232103 (2011).
\bibitem{correct} Here we have corrected the error in Eq.~(A9) in 
  Ref.~\onlinecite{Schliemann_ac_Rashba}. In addition, Eq.~(A6) in that paper should be
  corrected as $F_1=i\kappa(\kappa\sigma_z+\Lambda\sigma_x)/\sqrt{\kappa^2+\Lambda^2}$.
\end{thebibliography}
\end{document}